\documentclass[12pt,a4paper]{article}
\newcommand\blank[1]{}

\newcounter{tempeq}

\begin{document}
\begin{titlepage}
\vskip 0.5cm

\vskip 1.5cm
\begin{center}
{\Large{\bf
 On the generalized  unitary parasupersymmetry
 algebra of Beckers-Debergh
}}
\end{center}
\vskip 0.8cm
\centerline{A. Chenaghlou
\footnote{{\tt a.chenaghlou@sut.ac.ir}}}
\vskip 0.4cm
\centerline{\sl\small Physics Department,  Faculty of Science,
 Sahand University of Technology,  \,}
\centerline{\sl\small
 P O Box 51335-1996,  Tabriz,  Iran\,}
\vskip 1cm
\centerline{H. Fakhri
\footnote{{\tt hfakhri@theory.ipm.ac.ir}}}
\vskip 0.4cm
\centerline{\sl\small Institute for Studies
in Theoretical Physics and Mathematics (IPM), \,}
\centerline{\sl\small
 P O Box 19395-5531, Tehran, Iran\,}
\centerline{\sl\small and \,}
\centerline{\sl\small Department of Theoretical Physics and Astrophysics,
Physics Faculty,   \,}
\centerline{\sl\small
Tabriz University,  P O Box 51664, Tabriz, Iran \,}

\vskip 1.5cm

\begin{abstract}
\noindent
An appropriate  generalization of the unitary parasupersymmetry algebra of
Beckers-Debergh to  arbitrary   order is presented in this
paper. A special representation for realizing of the  even arbitrary order unitary
parasupersymmetry algebra of Beckers-Debergh is analyzed by one
dimensional shape invariance solvable models, 2D and
3D quantum solvable models obtained
from the shape invariance theory as well. In particular
in the special representation, it is  shown
that the isospectrum Hamiltonians
consist of   the two partner
 Hamiltonians of the shape invariance theory.
\end{abstract}
\end{titlepage}
\setcounter{footnote}{0} 
\def\thefootnote{\fnsymbol{footnote}}

\section{Introduction}

Supersymmetry (symmetry between the fermionic and bosonic degrees
of freedom) has important role \cite{reza1,reza2} in analyzing of
the quantum mechanical systems, since it can study remarkable
properties including the degeneracy structure of the energy
spectrum, the relations among the energy spectra  of the various
Hamiltonians and etc.. In particular in this theory, the energy
eigenvalues are necessarily non-negative  and the energy   of
non-zero (zero) ground state is related to the broken (unbroken)
supersymmetry. For the first time, Rubakov and Spiridonov
\cite{reza3} extended  the supersymmetry which is called
parasupersymmetry and it describes  an essential symmetry between
bosons and parafermions. Then, it was realized  that the
parasupersymmetry presented in Ref. \cite{reza3} is of order $p=2$
and  Khare \cite{reza4,reza5} generalized the Rubakov-Spiridonov (R-S)
parasupersymmetry to arbitrary order  $p \geq 1$:
\begin{eqnarray}
& &Q_{1}^{p}\,Q_{1}^{\dag} + Q_{1}^{p-1}\,Q_{1}^{\dag}\,Q_{1} + \cdots +
Q_{1}^{+}\,Q_{1}^{p}=2p\,Q_{1}^{p-1}\,H,
\nonumber\\
& &Q_{1}^{p+1}=0,
\nonumber\\
& &[H,Q_{1}]=0.
\end{eqnarray}
Here, $Q_{1}$ and $H$ stand for a parasupercharge and the bosonic
Hamiltonian.
 The relations (1)  for $p=1$ and $p=2$ are reduced to the
 supersymmetry algebra and the
Rubakov-Spiridonov (R-S)  parasupersymmetry algebra
\cite{reza3}, respectively. In addition, the relations (1)  describe  the unitary
parasupersymmetry  algebra of arbitrary order  $p$,
since $Q_{1}^{\dag}$ is the Hermitian conjugate  of the parasupercharge
$Q_{1}$.  So,  it is evident that  similar relations are satisfied
under interchange of $Q_{1}$ and
$Q_{1}^{\dag}$ in (1).

Before the extension of the R-S  parasupersymmetry  algebra to
arbitrary order $p$, for the $p=2$ case  remarkable and interesting
discussions
had been made [6-9].
For instance, the motion of
a spin-1 particle  along the $z$-axis in the presence of a magnetic
field can be described by a parasupersymmetric
 Hamiltonian which is obtained by the simple harmonic
 oscillator and Morse potentials \cite{reza10},  and also
 the problem was generalized \cite{reza11} for the spin-$\frac{p}{2}$
 particles. Meanwhile, in the context of quantum field theory, the
 R-S   parasupersymmetry  algebra of order $p=2$  leads
   to the infinite bosonic and parafermionic variables \cite{reza12}.
 In a special formulation of the R-S parasupersymmetric quantum mechanics
 \cite{reza13},  the Hamiltonian for which the energy spectrum cannot
 be negative is expressed in terms of an explicit function of the parasupercharge  $Q_{1}$. However,
 in general for the Rubakov-Spiridonov parasupersymmetric theory of
 arbitrary order the bosonic Hamiltonian cannot be obtained directly  in
 terms of the parasupercharge  $Q_{1}$,  and the energy eigenvalues are not
 necessarily non-negative. Moreover, there is no connection
 between the non-zero (zero) ground-state energy and the broken
 (unbroken) parasupersymmetry. Also, it has been shown
 \cite{reza5,reza14} that there are $p-1$ other conserved parasupercharges  and $p$
 bosonic constants in the R-S parasupersymmetric theory
 of arbitrary order  $p$.

In an early work, Infeld and Hull \cite{reza15} studied the
factorization and algebraic solutions of the bound state problems
and later,  Gendenshtein \textit{et.al.} considered the subject in
the framework of the shape invariance symmetry as an important
aspect of solvability  of wide range of the 1D quantum mechanical
models. It must be mentioned that the factorization and the shape
invariance symmetry have obtained a very helpful approach in the
representation of the supersymmetry theory [16-24]. Recently, most
of the one dimensional shape invariant solvable quantum mechanical
models have been classified into two bunches. The first bunch
\cite{reza25} includes models for which the shape invariance
parameter is the main  quantum number $n$. On the other hand, in
the second bunch \cite{reza11} the shape invariance  parameter of
the models is the secondary
 quantum number
$m$. Meanwhile, it has been shown in Ref. \cite{reza11}
 that the R-S parasupersymmetry   algebra
of arbitrary order $p$ is realized   by the shape invariant
quantum mechanical models so that the algebra can be represented
by the quantum mechanical states of the models. For realizing the algebra,
it has also been   shown that   the
bosonic Hamiltonian involves $p+1$ isospectrum   Hamiltonians.
 This fact has been studied in detail for the second
bunch of the shape invariant models in Ref. \cite{reza11}.
 In fact,   $p+1$ isospectrum Hamiltonians  are
obtained by adding $p$ appropriate constants to the factorized
Hamiltonians
 $\frac{1}{2}A(1)B(1)$, $\frac{1}{2}A(2)B(2)$, ..., $\frac{1}{2}A(p)B(p)$,
 and by adding the corresponding constant of the last Hamiltonian to its
partner Hamiltonian \textit{i.e.}  $\frac{1}{2} B(p)A(p)$ as well. Here, the
operators $B(l)$ and $A(l)$; $l=1,2,...,p$ are the raising and
lowering operators of the  quantum states of the
shape invariance theory,  respectively. One of the other successes  of
the R-S parasupersymmetry theory   of arbitrary
order  $p$ is the fact that it  can be realized [26-28]
 by the 2D and 3D solvable quantum
 mechanical models obtained from the shape invariance symmetry.

 In 1990 another formulation of  the unitary parasupersymmetry
 algebra of order $p=2$ was introduced by Beckers and Debergh \cite{reza29} as
 follows:
\begin{eqnarray}
& &[Q_{1},[Q_{1}^{\dag},Q_{1}]]=2\,Q_{1}\,H
\nonumber\\
& &[Q_{1}^{\dag},[Q_{1},Q_{1}^{\dag}]]=2\,Q_{1}^{\dag}\,H
\nonumber\\
& &Q_{1}^{3}={Q_{1}^{\dag}}^{3}=0
\nonumber\\
& &[H,Q_{1}]=[H,Q_{1}^{\dag}]=0.
\end{eqnarray}
 The appropriate
form of the parasupersymmetric Hamiltonian presented in the Eqs. (2)
has been constructed in Ref. \cite{reza29}  by using the generators of the
simple harmonic oscillator, and  the algebra (2)  has been
represented   by the quantum states of the simple harmonic
oscillator. But, a generalization of the Beckers-Debergh (B-D) parasupersymmetry  algebra to
arbitrary order   $p\geq2$ which is satisfied by the
  quantum mechanical systems has not been found
yet. In general, the R-S parasupersymmetry  algebra has
been much more successful than the B-D parasupersymmetry
algebra. Nevertheless, for the B-D parasupersymmetry algebra of order $p=2$, it has
been performed useful discussions.
 For example, Mostafazadeh has proved  that \cite{reza30} for both Rubakov-Spiridonov
 and Beckers-Debergh formulations of the  parasupersymmetric
 quantum mechanics of order $p=2$, the degeneracy structure of the energy
 spectrum can be derived using a thorough analysis of the
 parasupersymmetry  algebra. He showed that the result is independent of
 the details of the Hamiltonian, for example,  the degeneracy
 structure is not related to the dimension of the coordinate
 manifolds that the Hamiltonian is defined on it.
 Moreover, it has been shown that in general the Rubakov-Spiridonov
 (R-S)
  and Beckers-Debergh (B-D) systems possess identical degeneracy
  structures  \cite{reza30}. Also,  similar to
     the Witten index of the supersymmetric
   quantum mechanics, for the two kinds of ($p=2$)
  parasupersymmetry systems (B-D and R-S) a new set of topological
  invariants
   has been obtained \cite{reza30,reza31}.

In this paper it is intended to generalize the unitary
parasupersymmetry algebra of B-D. Also, we introduce a special
representation of the B-D unitary parasupersymmetry algebra of
even arbitrary order by 1D shape invariance solvable models,  and
some 2D and 3D quantum solvable models as well. Meanwhile, the
 B-D unitary parasupersymmetry algebra of even arbitrary order
 $p=2k$ is introduced with $2k$ independent conserved parasupercharges.

\section{Towards an appropriate generalization of the\,\,\,\,\,\,\,\,\,\,  B-D parasupersymmetry
algebra}

In order to  generalize the B-D
unitary parasupersymmetry algebra,
we shall take into account the parafermionic operators $b$ and
$b^{\dag}$ of arbitrary order $p$ which have been used by Khare
\cite{reza4,reza5}. In fact in moving from statistics to parastatistics, the parafermionic operators are
considered as the following $(p+1)\times(p+1)$ matrices:
\begin{equation}
(b)_{i,j}=C_{j}\delta_{i,j+1}
\hspace{.5in}
(b^{\dag})_{i,j}=C_{i}\delta_{i+1,j}
\hspace{.5in}i,j=1,2,...,p+1
\end{equation}
where the coefficients $C_{j}$ are given by
\begin{equation}
C_{j}:=\sqrt{j(p-j+1}=C_{p-j+1} \hspace{.5in}
j=1,2,...,p+1.
\end{equation}
It can be easily shown that
\begin{equation}
C_{1}C_{2}...C_{p}=p!
\end{equation}
 Using the parafermionic operators $b$ and $b^{\dag}$, one may
obtain a spin-$\frac{p}{2}$ representation for the group SU(2).
Indeed, by defining the operators $J_{\pm}$ and $J_{3}$ as:
\begin{eqnarray}
& &J_{+}:=b^{\dag} \hspace{1in}J_{-}:=b
\nonumber\\
&
&J_{3}:=\frac{1}{2}\,[b^{\dag},b]=diag\left(\frac{p}{2},\frac{p}{2}-1,...
,-\frac{p}{2}+1,-\frac{p}{2}\right),
\end{eqnarray}
 the following commutation relations corresponding to the Lie
algebra SU(2) may be derived
\begin{equation}
[J_{+},J_{-}]=2\,J_{3}\hspace{1in} [J_{3},J_{\pm}]=\pm \,J_{\pm}.
\end{equation}
It is noticed that the commutation relation of the parafermionic
operators $b$ and $b^{\dag}$ is proportional to the third
component of the spin-$\frac{p}{2}$ representation of the Lie group SU(2).

One of the well-known and important properties of the
parafermionic operators is \cite{reza4,reza5}
\begin{equation}
b^{p+1}={b^{\dag}}^{p+1}=0.
\end{equation}
Now it can be verified that the parafermionic operators $b$ and
$b^{\dag}$ of order $p$ possess the following new multilinear
structural relations ($p \geq 2$):

\setcounter{tempeq}{\value {equation}}
\renewcommand\theequation{\arabic{tempeq}\alph{equation}}
\setcounter{equation}{0}
\addtocounter{tempeq}{1}

\begin{equation}
\sum_{l=0}^{p} (-1)^{l+1} \,C_{l}^{p}\, b^{p-l}\, b^{\dag}\,
b^{l}=2\,\delta_{p,2}\,(-1)^p\,b^{p-1}
\end{equation}
\begin{equation}
\sum_{l=0}^{p} (-1)^{l+1} \,C_{l}^{p} \,{b^{\dag}}^{p-l}\, b\, {b^{\dag}}^{l}
=2\,\delta_{p,2}\,{b^{\dag}}^{p-1}
\end{equation}
where the constants $C_{l}^{p};\,\,l=0,\cdots,p$ are the Newton binomial
expansion coefficients. Using the Baker-Hausdorff formula for
two arbitrary operators $A$ and $B$ which is given by
\renewcommand\theequation{\arabic{equation}}
\setcounter{equation}{\value {tempeq}}
\begin{equation}
e^{\lambda A}\, B\, e^{-\lambda A} =
B+\frac{\lambda}{1!}[A, B]+\frac{\lambda^{2}}{2!}[A, [A,
B]]+\cdots,
\end{equation}
 the structural relations (9) between the operators $b$ and
$b^{\dag}$ may be written down as
\setcounter{tempeq}{\value {equation}}
\renewcommand\theequation{\arabic{tempeq}\alph{equation}}
\setcounter{equation}{0}
\addtocounter{tempeq}{1}

\begin{eqnarray}
   & &\underbrace{[b,[b,\cdots,[b},[b^{\dag},b]]\cdots]]=2\,\delta_{p,2}\,(-1)^p\,b^{p-1}
\nonumber\\
   & & (p-1)-\hbox{times}
\end{eqnarray}

\begin{eqnarray}
&
&\underbrace{[b^{\dag},[b^{\dag},\cdots,[b^{\dag}},[b,b^{\dag}]]\cdots]]
=2\,\delta_{p,2}\,{b^{\dag}}^{p-1}.
\nonumber\\
& &(p-1)-\hbox{times}
\end{eqnarray}
It is seen that the multilinear expressions in the left-hand sides
of the relations (11) are described in terms of the parafermionic
operators $b$ and $b^{\dag}$ in the right-hand sides. Actually,
the multilinear relations (11) between the parafermionic operators
$b$ and $b^{\dag}$ propose similar relations between the
parasupercharges of the parasupersymmetric quantum mechanics of
order $p$. The relations (8) and (11) indicate that there exist
the conserved parasupercharges $Q_{1}$ and $Q_{1}^{\dag}$ of order $p$,
and the bosonic Hamiltonian so that they generate the following
parasupersymmetry algebra:
\renewcommand\theequation{\arabic{equation}}
\setcounter{equation}{\value {tempeq}}

\setcounter{tempeq}{\value {equation}}
\renewcommand\theequation{\arabic{tempeq}\alph{equation}}
\setcounter{equation}{0}
\addtocounter{tempeq}{1}

   \begin{eqnarray}
   & &\underbrace{[Q_{1},[Q_{1},\cdots,[Q_{1}},[Q_{1}^{\dag},Q_{1}]]\cdots]]=2(-1)^p\,Q_{1}^{p-1}\,H
\nonumber\\
   & & (p-1)-\hbox{times}
\end{eqnarray}

\begin{eqnarray}
&
&\underbrace{[Q_{1}^{\dag},[Q_{1}^{\dag},\cdots,[Q_{1}^{\dag}},[Q_{1},Q_{1}^{\dag}]]\cdots]]=2\,{Q_{1}^{\dag}}^{p-1}\,H
\nonumber\\
& &(p-1)-\hbox{times}
\end{eqnarray}

\begin{equation}
Q_{1}^{p+1}={Q_{1}^{\dag}}^{p+1}=0
\end{equation}
\begin{equation}
[H,Q_{1}]=0
\end{equation}
\begin{equation}
[H,Q_{1}^{\dag}]=0.
\end{equation}
It can be easily verified that the relations (12) for any
arbitrary $p$ are closure under Hermitian conjugation. Meanwhile,
the algebraic relations (2) are a special case ($p=2$) of the
relations (12). Therefore, the relations (12) is an appropriate
generalization of the B-D unitary parasupersymmetry algebra with
the bosonic Hamiltonian $H$ and the parafermions of arbitrary
order $p$ i.e. $Q_{1}$ and $Q_{1}^{\dag}$.
\renewcommand\theequation{\arabic{equation}}
\setcounter{equation}{\value {tempeq}}

\section{A special representation for realizing of the even arbitrary
order unitary parasupersymmetry algebra of B-D by quantum solvable
models}
In this section we analyze a special representation for the B-D
quantum mechanical unitary parasupersymmetry algebra of even
arbitrary order $p=2k$ by wide range of the 1D, 2D and 3D solvable
models. In fact, the 1D models are the solvable models obtained
from the two approaches of the factorization with respect to the
main and secondary quantum numbers i.e. $n$ and $m$. On the other
hand, the 2D and 3D models representing the algebraic relations
(12) for $p=2k$ are some known quantum mechanical models on the
homogeneous manifolds $SL(2,c)/GL(1,c)$ and the group manifolds
$SL(2,c)$. Now in order to obtain the mentioned representations,
we only introduce two bunches of the shape invariance models which have
been classified before \cite{reza11,reza25}.

In master  function theory, a function $A(x)$ which is at most
of second order in terms of $x$,  and a non-negative weight function
$W(x)$ defined in an interval $(a,b)$ may be chosen so that
 $(1/W(x))(d/dx)(A(x)W(x))$ is a polynomial of at most first
 order. For a given master function $A(x)$ and its corresponding weight
 function $W(x)$, it has been shown that the eigenvalue equations of
   the  one dimensional partner  Hamiltonians
 corresponding to the first bunch of the superpotentials,  which are
 obtained from the factorization with respect to the main
 quantum number $n$,  will be \cite{reza25}
\begin{eqnarray}
&&B(n)A(n)\psi_{n}(\theta)=E(n)\psi_{n}(\theta) \nonumber \\
&&A(n)B(n)\psi_{n-1}(\theta)=E(n)\psi_{n-1}(\theta),
\end{eqnarray}
where the variable $\theta$ is  introduced by means of solving
the following first order differential equation
\begin{eqnarray}
d\theta=\frac{dx}{A(x)}.
\end{eqnarray}
The energy spectrum $E(n)$ and the wave function
$\psi_{n}(\theta)$ are given by

\begin{eqnarray}
&&E(n)=\frac{n}{4{\Big[}{\Big(}\frac{A(x)W^{\prime}(x)}{W(x)}
{\Big)}^{\prime}+nA^{\prime\prime}(x){\Big]}^2}{\Bigg \{}
4{{\Big(}\frac{A(x)W^{\prime}(x)}{W(x)}{\Big)}^{\prime}}^2
\nonumber\\
& &\hspace{12mm}\times
{\Big(}n{A^{\prime}}^2(0)-A(0){\Big(}\frac{A(x)W^{\prime}(x)}{W(x)}
{\Big)}^{\prime} {\Big )}
\nonumber \\
&&\hspace{12mm}-{\Big(}\frac{AW^{\prime}}{W}{\Big)}(0){\Big
(}A^{\prime\prime}(x){\Big(}\frac{AW^{\prime}}{W}{\Big)}(0)-
2A^{\prime}(0){\Big(}\frac{A(x)W^{\prime}(x)}{W(x)}{\Big)}^{\prime}{\Big
)}
\nonumber\\
& &\hspace{12mm}\times
{\Big (}2{\Big(}\frac{A(x)W^{\prime}(x)}{W(x)}{\Big)}^{\prime}+
nA^{\prime\prime}(x){\Big )} \nonumber \\ &&\hspace{12mm}
+n^2A^{\prime\prime}(0){\Big
(}{A^{\prime}}^2(0)-2A^{\prime\prime}(0)A(0){\Big )}{\Big
(}nA^{\prime\prime}(0)+
4{\Big(}\frac{A(x)W^{\prime}(x)}{W(x)}{\Big)}^{\prime}{\Big )}
\nonumber \\ &&\hspace{12mm}
-10nA(0)A^{\prime\prime}(x)
{{\Big(}\frac{A(x)W^{\prime}(x)}{W(x)}{\Big)}^{\prime}}^2{\Bigg\}},
\end{eqnarray}
\begin{eqnarray}
\psi_{n}(\theta)={\Big [}\frac{a_{n}}{\sqrt{W(x)}}{\Big (}\frac{d}{dx}{\Big )}^n{\Big
(}A^n(x)W(x){\Big )}{\Big ]}_{x=x(\theta)}.
\end{eqnarray}
Note that the prime symbol indicates the derivative with respect to
$x$.
 Moreover, the explicit forms of the raising and lowering operators
corresponding to the main  quantum number $n$ are,
respectively:
\begin{eqnarray}
& &B(n)=\frac{d}{d\theta}+ \frac{1}{2}{\Bigg[}nA^{\prime}(x)+
\frac{A(x)W^{\prime}(x)}{W(x)}
\Biggr.\nonumber\\
& &\hspace{1.25in}\Biggl.+n\frac{A^{\prime}(0){\Big(}\frac{A(x)W^{\prime}(x)}{W(x)}{\Big)}^{\prime}-
A^{\prime\prime}(x)
{\Big(}\frac{AW^{\prime}}{W}{\Big)}(0)}{{\Big(}\frac{A(x)W^{\prime}(x)}
{W(x)}{\Big)}^{\prime}+nA^{\prime\prime}(x)} {\Bigg]}_{x=x(\theta)} \nonumber \\
& &A(n)=-\frac{d}{d\theta}+ \frac{1}{2}{\Bigg[}nA^{\prime}(x)+
\frac{A(x)W^{\prime}(x)}{W(x)}
\Biggr.\nonumber\\
& &\hspace{1.35in}\Biggl.+n\frac{A^{\prime}(0){\Big(}\frac{A(x)W^{\prime}(x)}{W(x)}{\Big)}^{\prime}-
A^{\prime\prime}(x)
{\Big(}\frac{AW^{\prime}}{W}{\Big)}(0)}{{\Big(}\frac{A(x)W^{\prime}(x)}
{W(x)}{\Big)}^{\prime}+nA^{\prime\prime}(x)}{\Bigg]}_{x=x(\theta)}
\end{eqnarray}
The  change of variable $x=x(\theta)$  is substituted in the relations
(16) and (17) by  solving the first
order differential equation (14). Now, by choosing the suitable normalization
coefficients $a_{n}$ for the wave functions $\psi_{n}(\theta)$, one
may write down the shape invariance equations (13) as the
  raising and lowering relations:

\begin{eqnarray}
&&B(n)\psi_{n-1}(\theta)=\sqrt{E(n)}\psi_{n}(\theta) \nonumber \\
&&A(n)\psi_{n}(\theta)=\sqrt{E(n)}\psi_{n-1}(\theta).
\end{eqnarray}
The potentials like Coulomb, Rosen-Morse I, Rosen-Morse II and Eckart are
 involved in the first bunch of the solvable models.

 In master  function theory, the eigenvalue equations of the
  one dimensional partner Hamiltonian corresponding to the second
bunch of the superpotentials which are obtained from the
factorization with respect to the secondary quantum number
 $m$ are given by \cite{reza11}

\begin{eqnarray}
&&B(m)A(m)\psi_{n,m}(\theta)=E(n,m)\psi_{n,m}(\theta) \nonumber \\
&&A(m)B(m)\psi_{n,m-1}(\theta)=E(n,m)\psi_{n,m-1}(\theta).
\end{eqnarray}
In the factorization equations (19), the variable $\theta$ is
introduced
 by solving
the following first order differential equation

\begin{eqnarray}
d\theta=\frac{dx}{\sqrt{A(x)}}.
\end{eqnarray}
The energy spectrum and the eigenfunctions of the partner
Hamiltonians   obtained from  the factorization with respect to
the secondary quantum number $m$ are
\begin{eqnarray}
E(n,m)=-(n-m+1){\Big[}{\Big(}\frac{A(x)W^{\prime}(x)}{W(x)}{\Big)}^{\prime}
+\frac{1}{2}(n+m)A^{\prime\prime}(x){\Big]}
\end{eqnarray}
\begin{eqnarray}
\psi_{n,m}(\theta)={\Big [}\frac{a_{n,m}}{A^{(2m-1)/4}(x)W^{1/2}(x)}
{\Big(}\frac{d}{dx}
{\Big)}^{n-m}{\Big(}A^n(x)W(x){\Big)}{\Big]}_{x=x(\theta)},
\end{eqnarray}
where $m=0,1,2, \cdots ,n$. The explicit forms of the raising and lowering operators
 corresponding to  the secondary quantum number $m$ are given by
\begin{eqnarray}
&&B(m)=\frac{d}{d\theta}-
{\Big[}\frac{\frac{A(x)W^{\prime}(x)}{2W(x)}
+\frac{2m-1}{4}A^{\prime}(x)}{\sqrt{A(x)}}{\Big]}_{x=x(\theta)}
\nonumber \\
&&A(m)=-\frac{d}{d\theta}-{\Big[}\frac{\frac{A(x)W^{\prime}(x)}{2W(x)}
+\frac{2m-1}{4}A^{\prime}(x)}{\sqrt{A(x)}}{\Big]}_{x=x(\theta)}.
\end{eqnarray}
This time the change of variable $x=x(\theta)$ is
substituted in the Eqs.  (22) and (23) by solving the
first order differential equation (20). Similar to the Eqs.
(18), one may write down the raising and lowering relations
corresponding to the secondary quantum number $m$  on
the wave functions $\psi_{n,m}(\theta)$
 by  using  the factorization
equation (19) as:
\begin{eqnarray}
&&B(m)\psi_{n,m-1}(\theta)=\sqrt{E(n,m)}\psi_{n,m}(\theta) \nonumber \\
&&A(m)\psi_{n,m}(\theta)=\sqrt{E(n,m)}\psi_{n,m-1}(\theta).
\end{eqnarray}
The potentials like 3D harmonic oscillator, Scarf I, Scarf II,
Natanzon and generalized P\"oschl-Teller are included in the second bunch of the solvable models.
Moreover, simple harmonic oscillator and Morse potentials belong to
both of the solvable models.

   Now let us analyze a special  realization  of the
   B-D unitary parasupersymmetry  algebra of
   even arbitrary order $p=2k$ by means of
   the one dimensional quantum mechanical solvable
   models which are obtained from the factorization with respect to the
   secondary quantum number $m$. Similar procedure  can be made by
   means of the one dimensional quantum mechanical models which are
   obtained from  the shape invariance with respect to the main
   quantum number $n$. In order to realize  the B-D  unitary parasupersymmetry
   algebra of even arbitrary  order  $p=2k$, one may define the
   parafermionic generators $Q_{1}$ and $Q_{1}^{\dag}$ of order   $p=2k$,  and
   the bosonic operator $H$ as the following $(2k+1)\times(2k+1)$
   matrices:
\setcounter{tempeq}{\value {equation}}
\renewcommand\theequation{\arabic{tempeq}\alph{equation}}
\setcounter{equation}{0}
\addtocounter{tempeq}{1}

   \begin{eqnarray}
\left(Q_{1}\right)_{ll^{\prime}}:=\delta_{l+1, l^{\prime}}\, A(m)
,\hspace{1in}l=\hbox{odd}
\nonumber\\
\left(Q_{1}\right)_{ll^{\prime}}:=\delta_{l+1, l^{\prime}}\, B(m),
\hspace{1in}l=\hbox{even}
\end{eqnarray}

\begin{eqnarray}
\left(Q_{1}^{\dag}\right)_{ll^{\prime}}:=\delta_{l, l^{\prime}+1}\, A(m)
, \hspace{1in}l=\hbox{odd}
\nonumber\\
\left(Q_{1}^{\dag}\right)_{ll^{\prime}}:=\delta_{l, l^{\prime}+1}\,
B(m), \hspace{1in}l=\hbox{even}
\end{eqnarray}

\begin{equation}
\left(H\right)_{ll^{\prime}}:=\delta_{l, l^{\prime}}\, H_{l}
,\hspace{1in}l,l^{\prime}=1,2, \cdots,2k+1
\end{equation}
where $m=1, 2, \cdots, n$.
 It is evident that by choosing the definitions (25a) and (25b) for $Q_{1}$ and
 $Q_{1}^{\dag}$,  the relations (12c) are satisfied automatically.
  To satisfy  the Eqs.  (12a) and (12b), by using the definitions (25)
  in them   we obtain
\renewcommand\theequation{\arabic{equation}}
\setcounter{equation}{\value {tempeq}}

\setcounter{tempeq}{\value {equation}}
\renewcommand\theequation{\arabic{tempeq}\alph{equation}}
\setcounter{equation}{0}
\addtocounter{tempeq}{1}

\begin{eqnarray}
\left(\sum_{l=0}^{2k-1} (-1)^{l+1} \,C_{l}^{2k}\right)\,
\left( A(m)\,B(m)\right)^{k}\, A(m)= 2\left( A(m)\,B(m)\right)^{k-1}\, A(m)\,
H_{2k}
\nonumber\\
\left(\sum_{l=1}^{2k} (-1)^{l+1} \,C_{l}^{2k}\right)\,
\left( B(m)\,A(m)\right)^{k}\, B(m)= 2\left( B(m)\,A(m)\right)^{k-1}\, B(m)\,
H_{2k+1}
\end{eqnarray}

\begin{eqnarray}
\left(\sum_{l=1}^{2k} (-1)^{l+1} \,C_{l}^{2k}\right)\,
\left( B(m)\,A(m)\right)^{k}\, B(m)= 2\left( B(m)\,A(m)\right)^{k-1}\, B(m)\,
H_{1}
\nonumber\\
\left(\sum_{l=0}^{2k-1} (-1)^{l+1} \,C_{l}^{2k}\right)\,
\left( A(m)\,B(m)\right)^{k}\, A(m)= 2\left( A(m)\,B(m)\right)^{k-1}\, A(m)\,
H_{2}.
\end{eqnarray}
Considering the following identities
$$
\sum_{l=0}^{2k-1} (-1)^{l+1} \,C_{l}^{2k}=\sum_{l=1}^{2k} (-1)^{l+1}
\,C_{l}^{2k}=1,
$$
then the Eqs. (26)  give the following results
\renewcommand\theequation{\arabic{equation}}
\setcounter{equation}{\value {tempeq}}
\begin{eqnarray}
H_{1}=H_{2k+1}=\frac{1}{2}\,A(m)\,B(m)
\nonumber\\
H_{2}=H_{2k}=\frac{1}{2}\,B(m)\,A(m).
\end{eqnarray}
Meanwhile, by using the definitions (25),   the Eqs.  (12d) and (12e)  lead to
the following relations, respectively

\setcounter{tempeq}{\value {equation}}
\renewcommand\theequation{\arabic{tempeq}\alph{equation}}
\setcounter{equation}{0}
\addtocounter{tempeq}{1}

\begin{eqnarray}
& &A(m)\,H_{2l}=H_{2l-1}\,A(m)
\nonumber\\
& &B(m)\,H_{2l+1}=H_{2l}\,B(m), \hspace{1in} l=1, 2, \cdots, k
\end{eqnarray}

\begin{eqnarray}
& &A(m)\,H_{2l}=H_{2l+1}\,A(m)
\nonumber\\
& &B(m)\,H_{2l-1}=H_{2l}\,B(m), \hspace{1in} l=1, 2, \cdots, k.
\end{eqnarray}
It is noticed that the Eqs. (27) satisfy the relations (28a) and (28b),
and   additionally,   in order to determine the remaining components of the
bosonic Hamiltonian $H$ it is sufficient to substitute the relations (27)
 in the recursion Eqs.  (28a) and (28b).  Then,  one may obtain consistent
   solutions which are the same for the  Eqs.  (28a) and (28b)
as:

\renewcommand\theequation{\arabic{equation}}
\setcounter{equation}{\value {tempeq}}

\setcounter{tempeq}{\value {equation}}
\renewcommand\theequation{\arabic{tempeq}\alph{equation}}
\setcounter{equation}{0}
\addtocounter{tempeq}{1}

\begin{equation}
H_{2l-1}=\frac{1}{2}\,A(m)\,B(m), \hspace{1in} l=1, 2, \cdots, k+1
\end{equation}
\begin{equation}
H_{2l}=\frac{1}{2}\,B(m)\,A(m), \hspace{1.4in} l=1, 2, \cdots, k.
\end{equation}
The recent result declares  that the isospectrum   Hamiltonians
of the B-D  unitary parasupersymmetry  theory of even arbitrary order
$p=2k$
 are the two partner Hamiltonians $\Bigl(\frac{1}{2}A(m)B(m)((k+1)-\hbox{times})\,\,\,
 \hbox{and}\,\,\,
 \frac{1}{2}B(m)A(m) (k-\hbox{times}) \Bigr)$
 of the shape invariance theory with the energy spectrum
 $\frac{1}{2}E(n,m)$.
\renewcommand\theequation{\arabic{equation}}
\setcounter{equation}{\value {tempeq}}

Clearly,  the following $(2k+1)\times 1$ columns matrix
\begin{equation}
   \Psi(\theta)= \left( \begin{array}{lr} \psi_{n,m-1}(\theta) \\
                       \psi_{n,m}(\theta) \\ \psi_{n,m-1}(\theta)
 \\ \psi_{n,m}(\theta) \\ \,\,\,\,\,\,\,\vdots\\ \psi_{n,m-1}(\theta)
                           \\ \end{array}
             \right)_{(2k+1) \times 1}
\end{equation}
as  the basis represent the B-D unitary parasupersymmetry algebra of
even arbitrary order $p=2k$.  The
eigenvalue equation of the bosonic Hamiltonian $H$ is  written
down as
\begin{equation}
H\,\Psi(\theta)=E(n,m)\,\Psi(\theta).
\end{equation}
The representation of the parafermionic  generators $Q_{1}$ and $Q_{1}^{\dag}$
on the basis (30) by using the Eqs. (24) has  the following forms

\begin{equation}
  Q_{1}\, \Psi(\theta)=\sqrt{E(n,m)} \left( \begin{array}{lr} \psi_{n,m-1}(\theta) \\
                       \psi_{n,m}(\theta) \\ \psi_{n,m-1}(\theta)
 \\ \psi_{n,m}(\theta) \\ \,\,\,\,\,\,\,\vdots\\ \psi_{n,m}(\theta)\\\,\,\,\,\,\,\,0
                           \\ \end{array}
             \right) ,\,\,\,\,\,\,\,
 Q_{1}^{\dag}\, \Psi(\theta)=\sqrt{E(n,m)} \left( \begin{array}{lr}
  \,\,\,\,\,\,\,0\\ \psi_{n,m}(\theta) \\
                       \psi_{n,m-1}(\theta) \\ \psi_{n,m}(\theta)
 \\ \psi_{n,m-1}(\theta) \\ \,\,\,\,\,\,\,\vdots\\ \psi_{n,m-1}(\theta)
                           \\ \end{array}
             \right).
\end{equation}
It is easily seen that  the states
$Q_{1}^{l}\Psi(\theta)$ and ${Q_{1}^{\dag}}^{l}\Psi(\theta)$
 ($l=1, 2, \cdots, p=2k$) are eigenfunctions  of the bosonic Hamiltonian
 $H$.

If we consider the first bunch of the shape invariance  models, we
will be able to construct the parafermionic generators by means of
the raising and lowering operators $B(n)$ and $A(n)$, therefore
we can obtain the bosonic Hamiltonian $H$ including two
independent partner components $\frac{1}{2}A(n)B(n)$ and
$\frac{1}{2}B(n)A(n)$ with  the same energy spectrum
$\frac{1}{2}E(n)$. In  this case,  the basis of the
 representation of the B-D  unitary parasupersymmetry  algebra
 of even arbitrary order $p=2k$   is
 constructed by the eigenfunctions $\psi_{n}(\theta)$ and
$\psi_{n-1}(\theta)$.

In Ref.  \cite{reza27} the following eigenvalue equations have
been obtained by using the shape invariance  equations  (19)
\begin{eqnarray}
& &L_{+}\,L_{-}\,
\psi_{n,m}(\theta,\phi)=E(n,m)\,\psi_{n,m}(\theta,\phi),
\nonumber\\
& &L_{-}\,L_{+}\,
\psi_{n,m-1}(\theta,\phi)=E(n,m)\,\psi_{n,m-1}(\theta,\phi),
\end{eqnarray}
where the explicit differential forms of the operators $L_{+}$
and $L_{-}$ are given by
\begin{eqnarray}
& &L_{+}=e^{i\phi}\left(\frac{\partial}{\partial \theta}
+\frac{i}{2} \frac{A^{\prime}(x)}{\sqrt{A(x)}} \Big
|_{x=x(\theta)}\frac{\partial}{\partial \phi}
\right.\nonumber\\
& &\hspace{.9in}\left.-\left
[\frac{1}{2\sqrt{A(x)}}\left(\frac{A(x) W^{\prime}(x)}{W(x)}\right)
+\frac{1}{4\sqrt{A(x)}}A^{\prime}(x)\right]_{x=x(\theta)}\right)
\nonumber\\
\nonumber\\
& &L_{-}=e^{-i\phi}\left(-\frac{\partial}{\partial \theta}
+\frac{i}{2} \frac{A^{\prime}(x)}{\sqrt{A(x)}} \Big
|_{x=x(\theta)}\frac{\partial}{\partial \phi}
\right.\nonumber\\
& &\hspace{.9in}\left.-\left
[\frac{1}{2\sqrt{A(x)}}\left(\frac{A(x) W^{\prime}(x)}{W(x)}\right)
-\frac{1}{4\sqrt{A(x)}}A^{\prime}(x)\right]_{x=x(\theta)}\right).
\end{eqnarray}
Now taking into account the following operators
\begin{eqnarray}
& &L_{3}=-i\frac{\partial}{\partial\phi}
\nonumber\\
& &I=1
\end{eqnarray}
then, the operators $L_{+}$, $L_{-}$, $L_{3}$ and $I$ constitute
the Lie algebra $gl(2,c)$ i.e.
\begin{eqnarray}
& &\left[L_{+}, L_{-}\right]=-A^{\prime\prime}(x) L_{3}- \left(\frac{A(x)
W^{\prime}(x)}{W(x)}\right)^{\prime}I
\nonumber\\
& &\left[L_{3}, L_{\pm}\right]=\pm L_{\pm}
\nonumber\\
& &\left[\textbf{L}, I\right]=0.
\end{eqnarray}
The change of variable $x=x(\theta)$, which is used in the Eqs.
(34), is obtained by solving the differential Eq. (20). It has
been shown in Ref. \cite{reza27} that the Casimir of the generators
$L_{+}$, $L_{-}$, $L_{3}$ and $I$ is the corresponding Hamiltonian
of the charged particle on the homogeneous manifolds
$SL(2,c)/GL(1,c)$ in the presence of magnetic monopole with
degeneracy group $GL(2,c)$. The wave functions $\psi_{n,m}(\theta,\phi)$
, which represent the Lie algebra $gl(2,c)$ as
\begin{eqnarray}
& &L_{+}\,
\psi_{n,m-1}(\theta,\phi)=\sqrt{E(n,m)}\,\psi_{n,m}(\theta,\phi)
\nonumber\\
& &L_{-}\,
\psi_{n,m}(\theta,\phi)=\sqrt{E(n,m)}\,\psi_{n,m-1}(\theta,\phi)
\nonumber\\
& &L_{3}\,\psi_{n,m}(\theta,\phi)=m\,\psi_{n,m}(\theta,\phi)
\nonumber\\
& &I\,\psi_{n,m}(\theta,\phi)= \psi_{n,m}(\theta,\phi),
\end{eqnarray}
describe the two dimensional quantum states of the charged
particle on the homogeneous manifolds  $SL(2,c)/GL(1,c)$
in the presence of the magnetic monopole
and they are given by
\begin{equation}
\psi_{n,m}(\theta,\phi)= e^{im\phi}\, \psi_{n,m}(\theta).
\end{equation}

Now it can be easily shown that the quantum states
$\psi_{n,m}(\theta, \phi)$ also  represent the B-D unitary
 parasupersymmetry algebra of arbitrary order $p=2k$. In order to
 show the mentioned fact  it is sufficient to define the
 parafermionic generators $Q_{1}$ and $Q_{1}^{\dag}$ of order $p=2k$, and
 the bosonic operator $H$ as

\setcounter{tempeq}{\value {equation}}
\renewcommand\theequation{\arabic{tempeq}\alph{equation}}
\setcounter{equation}{0}
\addtocounter{tempeq}{1}

\begin{eqnarray}
& &\left(Q_{1}\right)_{ll^{\prime}}:=\delta_{l+1, l^{\prime}}\, L_{-}
,\hspace{1in}l=\hbox{odd}
\nonumber\\
&&\left(Q_{1}\right)_{ll^{\prime}}:=\delta_{l+1, l^{\prime}}\, L_{+},
\hspace{1in}l=\hbox{even}
\end{eqnarray}

\begin{eqnarray}
& &\left(Q_{1}^{\dag}\right)_{ll^{\prime}}:=\delta_{l, l^{\prime}+1}\,
L_{-}
, \hspace{1in}l=\hbox{odd}
\nonumber\\
&&\left(Q_{1}^{\dag}\right)_{ll^{\prime}}:=\delta_{l, l^{\prime}+1}\,
L_{+}, \hspace{1in}l=\hbox{even}
\end{eqnarray}

\begin{equation}
\hspace{1.7in}\left(H\right)_{ll^{\prime}}:=\delta_{l, l^{\prime}}\,
H_{l} ,\hspace{1.2in}l,l^{\prime}=1,2, \cdots,2k+1.
\end{equation}
Once again, the relations (39a) and (39b) satisfy the relation
(12c) automatically. Using the definitions (39a), (39b) and (39c),
the Eqs. (12a) and (12b) lead to the following results
\renewcommand\theequation{\arabic{equation}}
\setcounter{equation}{\value {tempeq}}

\begin{eqnarray}
& &H_{1}=H_{2k+1}=\frac{1}{2}\,L_{-}\,L_{+}
\nonumber\\
& &H_{2}=H_{2k}=\frac{1}{2}\,L_{+}\,L_{-}.
\end{eqnarray}
The Eqs. (12d) and (12e) by using the definitions (39), and the Eqs.
(40)  lead to
\begin{eqnarray}
& &H_{2l-1}=\frac{1}{2}\,L_{-}\,L_{+},\hspace{1in} l=1, 2, \cdots,
k+1
\nonumber\\
& &H_{2l}=\frac{1}{2}\,L_{+}\,L_{-}  \hspace{1in} l=1, 2, \cdots,
k.
\end{eqnarray}
Therefore, the operator $H$ of  the B-D unitary
parasupersymmetry algebra of arbitrary order $p=2k$ has two
Hamiltonian components
$\left(\frac{1}{2}\,L_{-}\,L_{+}((k+1)-\hbox{times})\,\,\, \hbox{and}
\,\,\,\frac{1}{2}\,L_{+}\,L_{-}\right.$
$\left.(k-\hbox{times}) \right)$
on the homogeneous manifolds
$SL(2,c)/GL(1,c)$ with the same energy spectrum $\frac{1}{2}
E(n,m)$. The representation basis of the parasupersymmetry algebra
in terms of the quantum states $\psi_{n,m}(\theta,\phi)$  which describe
the motion of the particle on the homogeneous manifolds
$SL(2,c)/GL(1,c)$  has the following form
\begin{equation}
   \Psi(\theta, \phi)= \left( \begin{array}{lr} \psi_{n,m-1}(\theta, \phi) \\
                       \psi_{n,m}(\theta, \phi) \\ \psi_{n,m-1}(\theta, \phi)
 \\ \psi_{n,m}(\theta, \phi) \\\,\,
 \,\,\,\,\,\,\vdots\\ \psi_{n,m-1}(\theta, \phi)
                           \\ \end{array}
             \right)_{(2k+1) \times 1}.
\end{equation}
The eigenvalue equation of the parasupersymmetric Hamiltonian is
\begin{equation}
H\,\Psi(\theta, \phi)= E(n,m)\,\Psi(\theta,\phi).
\end{equation}
The representation of the parafermionic generators $Q_{1}$ and
$Q_{1}^{\dag}$ by using the representation of the Lie algebra $gl(2,c)$
given in the relations (37) is  the relations
(32)  and the only difference is the fact that we must
substitute the quantum states $\psi_{n,m-1} (\theta,\phi)$ and
$\psi_{n,m} (\theta,\phi)$ instead of $\psi_{n,m-1} (\theta)$ and
$\psi_{n,m} (\theta)$,  respectively.

By choosing  the generators of the Lie algebra $gl(2,c)$
in terms of  three variables,  given in Ref. \cite{reza28},
we have  taken  into account the solvable quantum models on the
group manifolds $SL(2,c)$. In this case,  like the two dimensional
models, it can be also shown that the three dimensional solvable
quantum models on the group manifolds $SL(2,c)$ described in Ref. \cite{reza28}
represent the B-D unitary parasupersymmetry algebra of arbitrary
order $p=2k$.

\section{B-D unitary parasupersymmetry algebra with $p=2k$ conserved
parasupercharges}

For the B-D unitary parasupersymmetry algebra of even arbitrary order
$p=2k$ introduced by the relations (12), it can be shown that
 there are $2k-1$ independent conserved
parasupercharges in
addition to $Q_{1}$ . We denote these parasupercharges by $Q_{2},
Q_{3}, \cdots , Q_{2k}$. They and their Hermitian conjugates are
defined as (In this section, we only follow the discussion for the second bunch of
the 1D shape invariance models):

\setcounter{tempeq}{\value {equation}}
\renewcommand\theequation{\arabic{tempeq}\alph{equation}}
\setcounter{equation}{0}
\addtocounter{tempeq}{1}
\begin{eqnarray}
\left(Q_r\right)_{ll^{\prime}}&:=&\delta_{l+1, l^{\prime}}\, A(m)
\hspace{1in}r\neq l \hspace{1in}l=\hbox{odd}
\nonumber\\&:=&-\delta_{l+1, l^{\prime}}\, A(m)
\hspace{22mm}r=l \nonumber\\
\left(Q_r\right)_{ll^{\prime}}&:=&\delta_{l+1, l^{\prime}}\, B(m)
\hspace{1in}r\neq l \hspace{1in}l=\hbox{even}
\nonumber\\&:=&-\delta_{l+1, l^{\prime}}\, B(m)
\hspace{22mm}r=l
\end{eqnarray}
\begin{eqnarray}
\left(Q_r^{\dag}\right)_{ll^{\prime}}&:=&\delta_{l, l^{\prime}+1}\, A(m)
\hspace{1in}r\neq l^{\prime} \hspace{1in}l=\hbox{odd}
\nonumber\\&:=&-\delta_{l, l^{\prime}+1}\, A(m)
\hspace{22mm}r=l^{\prime} \nonumber\\
\left(Q_r^{\dag}\right)_{ll^{\prime}}&:=&\delta_{l, l^{\prime}+1}\, B(m)
\hspace{1in}r\neq l^{\prime} \hspace{1in}l=\hbox{even}
\nonumber\\&:=&-\delta_{l, l^{\prime}+1}\, B(m)
\hspace{22mm}r=l^{\prime}.
\end{eqnarray}
where $r=2, 3, \cdots , 2k$. In this section, $Q_{1}$ is considered as
the relations (25). Once more, the bosonic Hamiltonian $H$ is
defined as the relation (25c). The algebraic relations (12c) are
satisfied automatically for the definitions (44). The equations
(26a) and (26b) are also obtained from the equations (12a) and
(12b) for the parasupercharges (44a) and (44b), respectively.
Meanwhile, the algebraic relations (12d) and (12e) lead to the
equations (28) again. Therefore, the solutions (29) are obtained
for the bosonic Hamiltonian $H$ once again. Each of $2k$ conserved
parasupercharges along with a bosonic Hamiltonian $H$ satisfy
separately the parasupersymmetric algebraic relations. So,  we have
($r=1, 2, ..., 2k$):
\renewcommand\theequation{\arabic{equation}}
\setcounter{equation}{\value {tempeq}}

\setcounter{tempeq}{\value {equation}}
\renewcommand\theequation{\arabic{tempeq}\alph{equation}}
\setcounter{equation}{0}
\addtocounter{tempeq}{1}
\begin{eqnarray}
-C_0^{2k}Q_r^{2k}Q_r^{\dag}+C_1^{2k}Q_r^{2k-1}Q_r^{\dag}Q_r-\cdots
-C_{2k}^{2k}Q_r^{\dag}Q_r^{2k}=2Q_r^{2k-1}H
\end{eqnarray}
\begin{eqnarray}
-C_0^{2k}{Q_r^{\dag}}^{2k}Q_r+C_1^{2k}{Q_r^{\dag}}^{2k-1}Q_rQ_r^{\dag}-\cdots
-C_{2k}^{2k}Q_r{Q_r^{\dag}}^{2k}=2{Q_r^{\dag}}^{2k-1}H
\end{eqnarray}
\begin{eqnarray}
Q_r^{2k+1}={Q_r^{\dag}}^{2k+1}=0
\end{eqnarray}
\begin{eqnarray}
[H,Q_r]=0
\end{eqnarray}
\begin{eqnarray}
[H,Q_r^{\dag}]=0.
\end{eqnarray}
Similar situation is occurred in the R-S arbitrary order
parasupersymmetry algebra.
\renewcommand\theequation{\arabic{equation}}
\setcounter{equation}{\value {tempeq}}

In addition to the relations (45c), the parasupercharges $Q_{1},
Q_{2}, \cdots , Q_{2k}$ and their Hermitian conjugates satisfy a
generalised form of the relations (45c) which are given by

\setcounter{tempeq}{\value {equation}}
\renewcommand\theequation{\arabic{tempeq}\alph{equation}}
\setcounter{equation}{0}
\addtocounter{tempeq}{1}
\begin{eqnarray}
&&Q_1^{a_1}Q_2^{a_2}\cdots Q_{2k}^{a_{2k}}=0 \\
&&{Q_1^{\dag}}^{a_1}{Q_2^{\dag}}^{a_2}\cdots {Q_{2k}^{\dag}}^{a_{2k}}=0,
\end{eqnarray}
where $a_{1}+a_{2}+ \cdots +a_{2k}=2k+1$. Furthermore, there exist $(2k+1)$ independent bosonic constants as
\renewcommand\theequation{\arabic{equation}}
\setcounter{equation}{\value{tempeq}}
\begin{equation}
\begin{array}{llll}
&&\hspace{39mm}\left(I_1\right)_{ll^{\prime}}:=\delta_{l, l^{\prime}}\nonumber\\
&&\left.
\begin{array}{llll}
&&\hspace{30mm}\left(I_s\right)_{ll^{\prime}}:= \delta_{l, l^{\prime}}
\hspace{20mm}l\neq s \nonumber \\
&&\hspace{30mm}\hspace{11mm}:=-\delta_{l,
l^{\prime}}\hspace{16mm}l=s
\end{array}
\right\}
s=2,3,...2k+1
\end{array}
\end{equation}
which commute with the bosonic Hamiltonian $H$:

\begin{eqnarray}
[H,I_s]=0 \hspace{30mm} s=1,2,3,\cdots,2k+1.
\end{eqnarray}
It is also noticed that the commutation relations of the
parasupercharges and the bosonic constants are closure, that is,

\begin{eqnarray}
[I_s,Q_r]=\sum_{l=1}^{l=2k}d_lQ_l \hspace{20mm} s=1,2,3,\cdots,2k+1
 \hspace{3mm}\mbox{and} \hspace{3mm} r=1,2,3,\cdots,2k
\end{eqnarray}
where the coefficients $d_{l}$ are constants. Similar relations
exist for the Hermitian conjugate of the parasupercharges which
are obtained by taking Hermitian conjugate of the relations (49). The bosonic
constants $I_{s}$ and the parasupercharges $Q_{r}$ satisfy the
mixed multilinear  relations which are a generalisation of the
relations (45a) and (45b). For example, one may introduce the
B-D unitary parasupersymmetry algebra of order $p=2$ with two
parasupercharges and three bosonic constants:

\begin{eqnarray}
&&Q_1= \left( \begin{array}{ccc} 0& A(m) & 0
\\ 0 & 0 & B(m)   \\ 0 & 0 &  0
\end{array} \right)\hspace{12mm}
Q_1^{\dag}= \left( \begin{array}{ccc} 0& 0 & 0
\\ B(m) & 0 & 0   \\ 0 & A(m) &  0
\end{array} \right) \nonumber \\
&&Q_2= \left( \begin{array}{ccc} 0& A(m) & 0
\\ 0 & 0 & -B(m)   \\ 0 & 0 &  0
\end{array} \right)\hspace{9mm}
Q_2^{\dag}= \left(\begin{array}{ccc} 0& 0 & 0
\\ B(m) & 0 & 0   \\ 0 & -A(m) &  0
\end{array} \right) \nonumber \\
&&I_1= \left( \begin{array}{ccc} 1& 0 & 0
\\ 0 & 1 & 0   \\ 0 & 0 &  1
\end{array} \right)\hspace{10mm}
I_2= \left( \begin{array}{ccc} 1& 0 & 0
\\ 0 & -1 & 0   \\ 0 & 0 &  1
\end{array} \right)\hspace{10mm}
I_3= \left( \begin{array}{ccc} 1& 0 & 0
\\ 0 & 1 & 0   \\ 0 & 0 &  -1
\end{array} \right)\hspace{10mm}\nonumber\\
&&H=\frac{1}{2}\left( \begin{array}{ccc} A(m)B(m)& 0 & 0
\\ 0 & B(m)A(m) & 0   \\ 0 & 0 &  A(m)B(m)
\end{array} \right).
\end{eqnarray}
Clearly, the three generators $H, Q_{1}, Q_{1}^{\dag}$ and also
the three generators $H, Q_{2}, Q_{2}^{\dag}$ satisfy separately
the algebraic relations (45) with $k=1$. Moreover, we have

\begin{eqnarray}
&&Q_1^2Q_2=Q_1Q_2^2=0 \hspace{27mm} [I_1,Q_1]=[I_1,Q_2]=0 \nonumber \\
&&[I_2,Q_1]=2Q_2  \hspace{35mm} [I_2,Q_2]=2Q_1 \nonumber \\
&&[I_3,Q_1]=Q_1-Q_2  \hspace{27mm} [I_3,Q_2]=Q_2-Q_1 \nonumber \\
&&[I_s,I_{s^{\prime}}]=[I_s,H]=0  \hspace{5mm} s,s^{\prime}=1,2,3.
\end{eqnarray}
The mixed multilinear relations which are a generalisation of the
relation (45a) with $k=1$, are given by

\begin{eqnarray}
&&-C_0^{2}Q_r^{2}Q_{r^{\prime}}^{\dag}I_2+C_1^{2}Q_rQ_{r^{\prime}}^{\dag}Q_r -C_{2}^{2}
Q_{r^{\prime}}^{\dag}Q_r^{2}I_3=2Q_{r^{\prime}}H \hspace{11mm}r\neq r^{\prime} \nonumber \\
&&-C_0^{2}Q_{r^{\prime}}Q_{r}Q_{r}^{\dag}+C_1^{2}Q_rQ_{r}^{\dag}Q_{r^{\prime}} -C_{2}^{2}
Q_{r}^{\dag}Q_{r^{\prime}}Q_rI_3=2Q_{r^{\prime}}H \hspace{5mm}r\neq r^{\prime} \nonumber \\
&&-C_0^{2}Q_rQ_{r^{\prime}}Q_{r}^{\dag}I_2+C_1^{2}Q_{r^{\prime}}Q_{r}^{\dag}Q_{r} -C_{2}^{2}
Q_{r}^{\dag}Q_{r}Q_{r^{\prime}}=2Q_{r^{\prime}}H \hspace{5mm}r\neq r^{\prime}.
\end{eqnarray}
Note that when $r=r^{\prime}$, the multilinear relations are the
same as the relations (45a) with $k=1$. The generalised mixed
multilinear relations of the relations (45b) with $k=1$ are
obtained by taking Hermitian conjugate  of the relations (52).

\section{Conclusions}
In this paper  an appropriate generalization of the B-D
unitary parasupersymmetry algebra of arbitrary order $p$ is
presented.  It is   shown  that in a special approach
the generalization for even
arbitrary  order $p=2k$  can be represented by the one dimensional
shape invariance quantum models. In this approach, the partner Hamiltonians
of the shape invariance theory are the isospectrum components of
the bosonic Hamiltonian in the B-D unitary parasupersymmetric
theory. Also, as we mentioned before, the 2D and 3D quantum models on the homogeneous
manifolds $SL(2,c)/GL(1,c)$ and the group manifolds $SL(2,c)$
obtained from the shape invariance approach with respect to the
secondary quantum number $m$ realize the B-D unitary
parasupersymmetry algebra of arbitrary order $p=2k$. At the same
time, the bosonic Hamiltonian $H$ for the 2D and 3D models has two
isospectrum components. Meanwhile, the B-D unitary parasupersymmetry
algebra of even arbitrary order $p=2k$ is generalised to a B-D
unitary parasupersymmetry algebra in the presence of the $2k$ conserved
parasupercharges with the mixed multilinear relations.
\section*{Acknowledgment}

We would like to thank the
referee for fruitful comments which lead to a new section (section 4)
in our paper, and the  research council of Sahand University
of Technology for financial support.

\end{document}